# Bridging the Gap: the case for an 'Incompletely Theorized Agreement' on AI policy

Charlotte STIX[1,a*], Matthijs M. MAAS[2,b]

[a] Eindhoven University of Technology, The Netherlands.
[b] Centre for the Study of Existential Risk, University of Cambridge

**Abstract**: Recent progress in artificial intelligence (AI) raises a wide array of ethical and societal concerns. Accordingly, an appropriate policy approach is needed today. While there has been a wave of scholarship in this field, the research community at times appears divided amongst those who emphasize 'near-term' concerns, and those focusing on 'long-term' concerns and corresponding policy measures. In this paper, we seek to map and critically examine this alleged 'gulf', with a view to understanding the practical space for inter-community collaboration on AI policy. This culminates in a proposal to make use of the legal notion of an 'incompletely theorized agreement'. We propose that on certain issue areas, scholars working with near-term and long-term perspectives can converge and cooperate on selected mutually beneficial AI policy projects all the while maintaining divergent perspectives.

## 1. Introduction

The prevailing uncertainty around the trajectory and impact of artificial intelligence (AI) makes it clear that appropriate technology policy approaches are urgently needed. The possible negative ethical and societal impacts of AI are considerable: from algorithmic bias to AI-enabled surveillance, and from lethal autonomous weapons systems to widespread technology-induced unemployment. Moreover, some forecast that continuing progress in AI capabilities will eventually make AI systems a 'general-purpose technology' [1], or may even enable the development of 'high-level machine intelligence' (HLMI) [2] or other 'transformative' capabilities [3,4]. Debate on these latter scenarios is diverse, and has at times focused on what some have referred to as 'Artificial General Intelligence' (AGI) [5]. On the surface, those concerned with AI's impacts can appear divided between those who focus on discernible problems in the near-term, and those who focus on more uncertain problems in the longer-term [6–9].

[1] PhD Candidate, Philosophy and Ethics Group, Department of Industrial Engineering and Innovation Sciences, Eindhoven University of Technology, The Netherlands. Correspondence* should be addressed to c.stix@tue.nl. ORCID iD: 0000-0001-5562-9234. During the writing of this paper the author acted as Coordinator of the European Commission's High Level Expert Group on Artificial Intelligence. The author neither discloses confidential information nor makes use of information obtained during that work in this paper.
[2] Postdoctoral Research Associate, Centre for the Study of Existential Risk, University of Cambridge. During the writing of this paper, the author was funded under a PhD grant by the University of Copenhagen, at the Centre for International Law and Governance (Faculty of Law). ORCID iD: 0000-0002-6170-9393.

This paper wants to investigate the dynamics and debates between these two communities, with an eye to fostering policy effectiveness through cooperation. In doing so, this paper seeks to take up the recent call to 'bridge the near- and long-term challenges of AI' [9]. The focus is not on the relative urgency of existing algorithmic threats (such as e.g. facial recognition or algorithmic bias), nor on the relative plausibility of various advanced AI scenarios (such as e.g. HLMI or AGI), nor do we mean to suggest that a long-term perspective is solely focused on or concerned with AGI [10–12]. Rather, the paper proposes that even if some community divergence exists, each group's overarching intention to contribute to responsible and ethical AI policy[3] would benefit from cooperation within key domains in order to maximise policy effectiveness. The paper suggests that differences may be overstated, and proposes that even if one assumes deep differences, these are not practically insurmountable. Consequently, it argues that the principle of an 'incompletely theorized agreement', originally derived from constitutional law, provides both philosophical foundations and historical precedent for a form of cooperation between divergent communities that enables progress on urgent shared issues, without compromising on their respective goals.

The paper proceeds as follows: in Section 2, we provide a short rationale for our proposed intervention. We briefly lay out the landscape of AI policy concerns and the structure of the associated AI ethics and policy community. This is followed by a discussion, drawing on historical cases as well as the contemporary challenges facing AI policy scholars, of how fragmentation within an expert community might hinder progress on key and urgent policies. In Section 3, we explore potential sources which could contribute to community divergence. We touch on epistemic and methodological disagreements and normative disagreements, and home in on pragmatic disagreements around the tractability of formulating AI policy actions today which maintain long-term relevance. We briefly review how serious these disagreements are, arguing that these trade-offs are often exaggerated, and do not need to preclude collaboration. Finally, in Section 4, we propose that one consolidating avenue to harness mutually beneficial cooperation for the purpose of effective AI policy could be anchored in the constitutional law principle of an 'incompletely theorized agreement'. This proposal works under the assumption that the influence of a community on policy making is significantly stronger if they act as a united front, rather than as manifold subgroups.

[3] We define 'AI policy' broadly, as concrete soft or hard governance measures which may take a range of forms such as principles, codes of conduct, standards, innovation and economic policy or legislative approaches, along with underlying research agendas, to shape AI in a responsible, ethical and robust manner. Our paper works under the assumption that policy making can positively influence the development and deployment of AI technology.

## 2. AI policy: A house divided?

Recent progress in AI has given rise to an array of ethical and societal concerns.[4] Accordingly, there have been calls for appropriate policy measures to address these. As an "omni-use technology" [13] AI has both potential for good [14–16] as well as for bad [17–19] applications. The latter include: various forms of pervasive algorithmic bias [20,21], challenges around transparency and explainability [22,23]; the safety of autonomous vehicles and other cyber-physical systems [24], or the potential of AI systems to be used in (or be susceptible to) malicious or criminal attacks [25–27]; the erosion of democracy through e.g. 'computational propaganda' or 'deep fakes' [28–30], and an array of threats to the full range of human rights [31,32]. The latter may eventually cumulate in the possible erosion of the global legal order by the comparative empowerment of authoritarian states [33,34]. Finally, some express concern that continued technological progress might eventually result in increasingly more 'transformative' AI capabilities [3], up to and including AGI. Indeed, a number of AI researchers expect some variation of 'high-level machine intelligence' to be achieved within the next five decades [2]. Some have suggested that, if those transformative capabilities are not handled with responsibility and care, such developments could well result in new and potential catastrophic risks to the welfare, autonomy, or even long-term survival of societies [35,36].

Looking at the current debate and scholarship involved in the aforementioned areas, we note, along with other scholars [6,8,37], that there appears to be a fuzzy split, along a temporal 'near-term'/'long-term' axis. This perception matters because, as in many other fields and contexts, a perceived or experienced distinction may eventually become a self-fulfilling prophecy [38]. This holds true even if the perceived differences are based on misperceptions or undue simplification by popular-scientific media [39–41]. Of course, fragmentation between near- and longer-term considerations of AI's impact is only one way to explore the growing community, and it may not be the sole issue to overcome in order to maximise policy impact. However, for the purpose of this paper our focus is on this specific gap.

### The policy advantages of collaboration: lessons from history

The current AI ethics and policy community is a young one. Policy shifts, on the other hand, take time. As such, it is difficult to clearly outline what impact current dynamics have, we are, after all, still in the early stages of these developments. Nevertheless, historical examples of adjacent fields can help to demonstrate and forecast how fragmentation, or, conversely cooperation, on policy goals within the AI ethics and policy community could strengthen impact on technology policy.

Why should potential fragmentation along an axis such as near- and longer-term concerns worry us? History shows that the structure of a field or community affects the ability of its members to shape and influence policy. Importantly, it shows that there is significant benefit derived from collaboration. We put forward three historic examples of adjacent fields to AI, meaning those that tackled equally new and emerging technologies. We briefly highlight one case where fragmentation may have contributed to a negative impact on the overall policy impact of the community and two cases where a collaborative effort yielded a positive impact on policy formulation.

[4] This paper perceives AI's ethical and societal concerns to be closely intertwined, and as such refers to the broader set of these actual and potential concerns throughout.

*Nanotechnology:* One community that arguably suffered from a public pursuit of fractious division was the nanotechnology community in the early 2000s [42,43]. Internal disagreements came to a head in the 2003 'Drexler-Smalley' debate [44], which cemented an oversimplified caricature of the field. In reviewing this case, it has been argued that 'para-scientific' media created "polarizing controversy that attracted audiences and influenced policy and scientific research agendas. [...] bounding nanotechnology as a field-in-tension by structuring irreconcilable dichotomies out of an ambiguous set of uncertainties." [38]. This showcases a missed opportunity within a fragmented community to come together in order to promote greater political engagement with the responsible development of the technology.

*Recombinant DNA:* In the 1970's, concerns arose over recombinant DNA (rDNA) technology. In particular, the ethical implications of the ability to reshape life, as well as fears over potential biohazards from new infectious diseases led the biotechnology community to come together at the 1975 Asilomar Conference on Recombinant DNA to set shared standards [45]. The conference is widely considered a landmark in the field [46]: the scientist's and lawyers' commitment to a forthright open and public discussion has been argued to have stimulated both public interest and grounded policymaker discussion about the social, political and environmental issues related to genetic biotechnology in medicine and agriculture [47].

*Ballistic missile defense arms control:* In the wake of the creation of the atom bomb, a number of scientists expressed dismay and sought to institutionalize global control of these weapons. Early efforts to this end, such as the 1946 Baruch Plan, proved unsuccessful [48,49]. However, by the 1950s-1960s, a new 'epistemic community' emerged, bringing together both technical and social scientists, specifically in opposition to the development of anti-ballistic missile (ABM) systems. This community proved able to develop and disseminate this new understanding of nuclear deterrence dynamics to policymakers [50]. They achieved this by maintaining a high level of consensus on concrete policy goals, by framing public discourse on the ethical goals, and by fostering links to both policymakers as well as to Soviet scientists. This allowed them to persuade key administration figures and shift policymaker norms and perceptions at home and internationally. Ultimately, setting the stage for the 1972 ABM Treaty, the first arms control agreement of this kind [50,51].

**The pitfalls of fragmented efforts in AI Policy**

While some of the historical context is surely different, a number of these historical dynamics may well transfer to the emerging AI ethics and policy community [52]. Those concerned with AI policy could benefit from exploring such historical lessons. This should be done with urgency.

Firstly, there is a *closing window of opportunity*. The field of AI policy is a relatively new one, which offers a degree of flexibility in terms of problem framings, governance instrument choice and design, and community alignment. Going forwards, however, this field has a high likelihood of becoming progressively more rigid as framings, public perceptions, and stakeholder interests crystallize. Current dynamics could therefore have far-reaching impacts, given the potential to lock in a range of path-dependencies, for example through particular framings of the issues at hand [53]. In this context, a divided community which potentially treats policymakers or public attention as a zero-sum good for competing policy projects, may compromise the legitimacy of its individual efforts in front of these. This could undercut the leverage of policy initiatives today and in the future. Worse, public quarrels or contestation may 'poison the well'. Policymakers may begin to perceive and treat a divided community as a series of interest groups rather than an 'epistemic community' with a multi-faceted but coherent agenda for beneficial societal impact of AI. Finally, from a policy perspective, it is important to note that while current regulatory initiatives are not always directly transferable, neither are they categorically irrelevant. As such, they can often provide the second-best tools for rapidly confronting new AI challenges. This has its own pitfalls, but is often superior to waiting out the slow and reactive formulation of new policies.

Moreover, the *risks are concrete and timely*. It is plausible that political moods will shift within the coming years and decades, in ways that make policy progress much harder. Furthermore, it is possible that other epistemic communities may converge and mobilize faster to embed and institutionalize alternative, less broadly beneficial framings of AI. Indeed, public and global framings of AI in recent years have seemed to drift towards narratives of competition and 'arms races' [54–56, but see also 57]. An inflection point for how societies use and relate to AI may eventually be reached. Missing such a window of opportunity could mean that the relative influence of those concerned with making the impact of AI beneficial (whether in the near-, or longer-term) will decline, right as their voices are needed most. Conversely, many gains secured today could have lasting benefits down the road.

## 3. Examining potential grounds for divergence

There are a range of factors that could contribute to clustering into fuzzy 'near-' and 'long-term' communities and different scholars may hold distinct and overlapping sets of beliefs on them [cf. 8]. In the following paragraphs, we provide our first attempt at mapping some of these factors.[5]

Some part of the divergence may be due to varying *epistemic* or *methodological* commitments. These could reflect varying levels of tolerance regarding scientific uncertainty and distinct views on the threshold of probability required before far-reaching action or further investigation is warranted. This means that concerns surrounding AI may depend on qualitatively different conceptions of 'acceptable uncertainty' for each group of observers. This may well be hard to resolve. Moreover, epistemic differences over the implicit or explicit disagreements of the modal standards in these debates, for example, debates over what types of data or arguments are admissible in establishing or contesting the plausibility or probability of risk from AI may contribute to further divergence. This could even lead to differential interpretations of evidence that are available. For instance, do empirically observed failure modes of present-day architectures [58–61] provide small-scale proof-of-concepts of the type of difficulties we might one day encounter in AI 'value alignment', or are such extrapolations unwarranted?

For our purposes, however, the most salient factor may be essentially *pragmatic*. Different perceptions of the empirical dynamics and path-dependencies of governing AI can inform distinct theories-of-change. These are intertwined with one's expectations about the tractability and relevance of formulating useful and resilient policy action today. In this context, Prunkl &

[5] It should be emphasized that this mapping is only an indicative sketch, and would be much enriched by further examination, for example through structured interviews or comprehensive opinion surveys.

Whittlestone [8] have recently argued that a more accurate picture and more productive dialogue could be achieved if scholars differentiated amongst the four dimensions on which views vary, in terms of the capabilities, impacts, certainty or extremity of AI systems. They emphasize that views on each of these questions fall on a spectrum. Taking this point on board, there are additional ways to cash out possible divergences. One debate might concern the question, *how long-lasting are the consequences of near-term AI issues?* If those that care about the longer-term are convinced that these issues will not have long-lasting consequences, or that they would eventually be swamped by the much larger trends and issues [3], then this could lead them to discount work on near-term AI problems. However, it is important to note that near-term issues are likely to considerably affect the degree to which society is vulnerable to longer-term dangers posed by future advanced AI systems. Short-term or medium-term issues [7,37] can easily increase society's general turbulence [62], or lock in counterproductive framings of AI or our relation to it. In general, we might expect many nominally near-term effects of AI on society (such as in surveillance; job automation; military capabilities) to scale up and become more disruptive as AI capabilities gradually increase [18,37]. Indeed, some longer-term scholars have argued that advanced AI capabilities considerably below the level of HLMI might already suffice to achieve a prepotence which could pose catastrophic risks [10]. This would make mid-term impacts particularly important to handle, and collaboration between different groups on at least some generalizable projects crucial.

Another pragmatic question or concern is over *how much leverage we have today to meaningfully shape policies* that will be applicable or relevant in the long term, especially if AI architectures or the broader political and regulatory environment change a lot in the meantime [8]. Some scholars may hold that future AI systems will be technically so different from today's AI architectures that research into this question undertaken today will not be relevant. Conversely, they might hold that such advanced AI capabilities may be so remote that the regulatory environment will have changed too much for meaningful policy work to be conducted right now [63]. These people might argue that we had better wait until things are clearer and we are in a better position to understand whether and what research is needed or meaningful.

In practice, this critique does not appear to be a very common or deeply held position. Indeed, as a trade-off it may be overstated. It is plausible that there are diverse areas on which both communities can undertake valuable research today, because the shelf life of current policy and research efforts might be longer than is assumed. To be sure, there is still significant uncertainty over whether current AI approaches can at all be scaled up to very advanced performance [64–66]. Nonetheless, research could certainly depart from a range of areas of overlap [67] and shared areas of concern [68,69].

Moreover, policy-making is informed by a variety of aspects which range across different time spans. Starting with political agendas that often reflect the current status quo, policy making is equally shaped by shifting public discourse, societal dynamics and high-impact shocks. The latter factor has played a key role in AI policy, where high-profile incidents involving algorithmic discrimination, lack of transparency, or surveillance have driven policy shifts, as seen for example in the pushback on discriminatory algorithms used in the UK visa selection processes [70], the Artificial Intelligence Video Interview Act regulating the use of AI in employee interviews, or the California B.O.T. Law requiring AI systems to self-identify [71,72].

In sum, it is plausible that many perceived 'barriers' to inter-community cooperation on policy are not all that strong, and that many 'tradeoffs' are likewise overemphasized. However, does that mean there are also positive, mutually productive opportunities for both communities to work on with regard to policy in spite of outstanding disagreements? What would such an agreement look like?

## 4. Towards 'incompletely theorized agreements' for AI policy

Above, we have reviewed potential sources for divergence within the community. We will now discuss how even in the context of apparent disagreement, pragmatic agreements on shared policy goals and norms could be reached.

We propose to adopt and adapt the legal principle of an 'incompletely theorized agreement' for this purpose. Legal scholarship in constitutional law and regulation has long theorized the legal, organizational and societal importance of such incompletely theorized agreements. Their key use is that they allow a given community to bypass or suspend [73,74] any theoretical disagreement on matters where (1) the disagreement appears relatively intractable and (2) there is an urgent need to address certain shared practical issues. Disagreements are intractable in cases where either it simply does not appear as if the question will be decisively resolved one way or the other in the near term, or where time and capacity to reason through all underlying disagreements is *limited* [73]. Incompletely theorized agreements can therefore apply to deep philosophical and ethical questions as much as to contexts of pervasive scientific uncertainty. The latter is especially the case on questions where it still remains unclear where and how we might procure the information that allows definitive resolution.

Incompletely theorized agreements are a fundamental component to well-functioning legal systems, societies, and organizations. They allow for stability and flexibility to get urgent things done [75]. These agreements have long played a key role in constitutional and administrative law, and have made possible numerous landmark achievements of global governance, such as the establishment of the Universal Declaration of Human Rights [75,76]. The framework has also been extended to other domains, such as the collective development and analysis of health-care policies in the face of pluralism and conflicting views [77].

Incompletely theorized agreements have broad similarities with the notion of an overlapping consensus, developed by John Rawls, which refers to the way adherents of different (and apparently inconsistent) normative doctrines can nonetheless converge on particular principles of justice to underwrite the shared political community [78]. This concept has been read as a key mechanism in the field of bioethics, serving to enable agreement despite different fundamental outlooks [79]. It also already plays a role in the existing literature on computer ethics [80], as well as in the field of intercultural information ethics [81]. Indeed, overlapping consensus has been proposed as a mechanism on which to ground global cooperation on AI policy across cultural lines [82].

If overlapping consensus can ground inter-cultural cooperation, incompletely theorized agreements might serve as a similar foundation for practical cooperation between near- and long-term perspectives. In a related context, Baum has suggested that policy interventions aimed at securing long-term resilience to various catastrophes can often involve significant co-benefits in the near-term, and so do not narrowly depend on all parties agreeing on the deep reasons for the policies proposed [83]. Could incompletely theorized agreements ground cooperation amongst AI policy communities? We suggest that they could.

**Incompletely theorized agreements in AI policy: examples and sketches**

There are a range of issue areas where both groups could likely locate joint questions they would want addressed, and shared goals for which particular AI policies should be implemented. This holds even if their underlying reasons for pursuing these are not fully aligned. Without intending to provide an exhaustive, in-depth or definitive overview, a brief survey might highlight various areas for cooperation.

For one, *gaining insight into- and leverage on the general levers of policy formation around AI* [52] is a key priority. Respectively, understanding what are the steps in the policymaking process which determine what issues get raised to political agendas and eventually acted upon, and which might be derailed by other coalitions [84]? Given the above, research into underlying social and societal developments is fruitful in order to advance all groups' ability to navigate mutually agreeable policy goals across the policy making cycle [85]. For example, research into when, where or why global AI governance institutions might become vulnerable to regulatory capture or institutional path dependency ought to be an important consideration, whatever one's AI concerns are [86,87].

On a more operational level, this can feed into joint investigation into the relative efficacy of various policy levers for AI governance. For example, insights into when and how AI research labs or individual researchers adopt, or alternately cut corners on, responsible and accountable AI, or the incentivization of shifts in workplace culture or employee norms, could shape the policy proposals the community might make. The question of how to promote prosocial norms in AI research environments is of core interest to both communities with an eye to technology policy [88]. This might cover e.g. whether to publicly name problematic performance (e.g. biased results; lack of safety) in commercial AI products results in tech companies actually correcting the systems [89]; or whether codes of ethics are effective at changing programmers' decision making on the working floor [90,91]. All of these could be fruitful areas of collaboration on eventual policy proposals for either community.

More specifically, there are a range of particular AI policy programs that we expect could be the site of an incompletely theorized agreement.

1. Incompletely theorized agreements could shape norms and policy debates over the *appropriate scientific culture for considering the impact and dissemination of AI research* [92,93], especially where it concerns AI applications with potentially salient misuses. The underlying reasons for such policies might differ. Some may be concerned over abuses of vulnerable populations, new vectors for criminal exploitation or the implications of new language models for misinformation [94,95]; and others over the long-term risks from the eventual open development of advanced AI systems [96]. Accordingly, incompletely theorized agreements in this area could converge on policies to shape researcher norms around improved precaution or reflection around the impact or potential misuse of research [97].

2. Another example might be found in the domain of the *global regulation of military uses of AI*. This area has already seen years of shared efforts and even collaboration amongst a coalition of activists, lawyers, and institutions departing from both a near-term as well as longer-term perspective, such as the Future of Life Institute [52].
3. Incompletely theorized agreements could ground productive policy cooperation on policy interventions aimed at *preserving the integrity of public discourse and informed decision-making in the face of AI systems*. Policies aimed at combating AI-enabled disinformation would be a natural site for incompletely theorized collaboration, because a society's epistemic security [98] is relevant from both a near-term and long-term perspective alike.
4. Similarly, incompletely theorized agreements surrounding the promotion of policies aimed at *securing citizens' (political) autonomy and independence from unaccountable perception control* could be promising. After all, practices of opaque technological management [99] or perception control [100] can enable authorities to increasingly shape individuals' and societies' behaviour and values. Regulation to restrict the deployment of such tools, to facilitate privacy-preserving AI, or to ensure transparency and accountability of the principals of such tools, are important from a near-term perspective concerned with the role of hyper nudges [101], algocracy [102], or surveillance capitalism [103]. Moreover, such policies are also critical to avert long-term worries over a value lock-in, whereby one generation might someday "invent a technology that will enable the agents alive at that time to maintain their values indefinitely into the future, controlling the broad sweep of the entire rest of the future of civilisation" [104].

Although the underlying motives for each group to pursue policies on the abovementioned domains may be partially distinct, these technical differences are arguably thwarted by the benefits derived from achieving impactful and effective policy measures together. In many of these cases, the practical benefits of an incompletely theorized agreement would be at least four-fold (1) to reduce public confusion around these topics; (2) to present policymakers with an epistemic community delivering integrated policy proposals; (3) to support the articulation of regulations or governance instruments for specific policy problems, which need not assume further advances in AI capabilities, but which are also not reliant on provisions or assumptions that are vulnerable to 'obsolescence' if or when such advances do occur [105–107], giving such policies a longer shelf-life; (4) to improve engagement of particular AI policies with the steadily increasing cross-domain nature of AI, which could help inform regulatory responses across domains. This is especially relevant because different fields (such as content moderation, healthcare, or the military) often confront different yet similar versions of underlying problems [108].

**Limitations of incompletely theorized agreements**

We do not wish to suggest that incompletely theorized agreements are an unambiguously valuable tool across all AI policy cases, or even a definite solution for any one policy case. Such agreements do suffer from a number of potential drawbacks or trade-offs, which both communities should consider before invoking them in any particular case.[6]

Firstly, depending on one's assumptions around the expected degree of change in AI or in its societal impacts, incompletely theorized agreements could prove brittle. Incompletely theorized agreements bypass a full examination of the underlying disagreements in order to facilitate pragmatic and swift action on particular policies on which both communities find themselves in practical alignment in a specific moment in time. This may create a lack of clarity over the boundary conditions of that practical agreement, along with opacity over whether, or where (i.e. for which future particular questions around AI policy) the practical agreement might suddenly break down for either or all parties.

Secondly, an incompletely theorized agreement is, in an important sense, a 'stopgap' measure more than a general ideal, permanent one. As discussed above, an incompletely theorized agreement might be most suited to situations where (a) practical policy action is urgently needed, and (b) underlying theoretical agreement by stakeholders on all engaged principles or questions does not seem close. However, over longer times, deeper inquiry and debate do appear necessary [73]. And there is always the possibility that agreement was not, in fact, intractable within the near term. As such, a premature leap by the community into an incompletely theorized agreement in order to achieve some policy X might inadvertently curb the very conversations amongst the communities which could have led both to eventually prefer policy Y instead, had their conversation been allowed to run its course.

Moreover, there is a key related point here, on which we should reflect. By advocating for the adoption of incompletely theorized agreements on AI policy today, we ourselves are in a sense assuming or importing an implicit judgment about the urgency of AI issues today, and about the intractability of the underlying debates. Yet these are two positions which others might contest. For example, by arguing that 'AI issues do not today meet that threshold of urgency that the use of an incompletely theorized agreement is warranted'. We wish to make this assumption explicit. At the same time, we expect that it is an assumption widely shared by many scholars working on AI policy, many of whom may well share a sense that the coming years will be a sensitive and even critical time for AI policies.

Thirdly, a sloppily formulated incompletely theorized agreement on an AI policy issue may not actually reflect convergence on particular policies (e.g. 'certification scheme for AI products with safety tests X, Y, Z'). Instead it might solidify on apparent agreement on vague mid-level principles or values (e.g. 'AI developers should ensure responsible AI development'). These may be so broad that they do not ground clear action at the level of actual policies. If this were to happen, incompletely theorized agreements might merely risk contributing to the already-abundant proliferation of broad AI principles or ethical frameworks on AI that have little direct policy impact. While the ecosystem of AI codes of ethics issued in recent years have certainly shown some convergence [109–111], they have been critiqued as being hard to operationalize, and for providing only the appearance of agreement, while masking underlying tensions in the principles' interpretation, operationalization, or practical requirements [93,112]. Situations where an incompletely theorized agreement does not manage to root itself at the level of concrete policies but only mid-level principles would be a worst-of-both-worlds scenario: it would reduce the ability of actors to openly reflect upon and resolve inconsistencies amongst- or disagreements about high-level principles, while not even affording improvements at facilitating concrete policies or actions in particular AI domains. To mitigate this risk, incompletely theorized agreements should therefore remain closely grounded in concrete and clearly actionable policy goals or outputs.

[6] We thank one reviewer for prompting this following discussion of the drawbacks of incompletely theorized agreements.

Nonetheless, while limitations such as these should be considered in greater detail, we argue that they do not categorically erode the case for implementing, or at least further examining the promise of this principle and tool for advancing responsible AI policy.

## 5. Conclusion

AI has raised multiple societal and ethical concerns. This highlights the urgent need for suitable and impactful policy measures in response. Nonetheless, there is at present an experienced fragmentation in the responsible AI policy community, amongst clusters of scholars focussing on ‘near-term’ AI risks, and those focussing on ‘longer-term’ risks. This paper has sought to map the practical space for inter-community collaboration, with a view towards the practical development of AI policy.

As such, we briefly provided a rationale for such collaboration, by reviewing historical cases of scientific community conflict or collaboration, as well as the contemporary challenges facing AI policy. We argued that fragmentation within a given community can hinder progress on key and urgent policies. Consequently, we reviewed a number of potential (epistemic, normative or pragmatic) sources of disagreement in the AI ethics community, and argued that these trade-offs are often exaggerated, and at any rate do not need to preclude collaboration. On this basis, we presented the novel proposal for drawing on the constitutional law principle of an ‘incompletely theorized agreement’, in order for the communities to set aside or suspend these and other disagreements for the purpose of achieving higher-order AI policy goals of both communities in selected areas. We therefore non-exhaustively discussed a number of promising shared AI policy areas which could serve as the sites for such agreements, while also discussing some of the overall limits of this framework.

This paper does not suggest that communities should fully merge or ignore differences whatever their source may be. To be sure, some policy projects will be relevant to one group within the community but not the other. Indeed, community heterogeneity and diversity is generally a good thing for a scientific paradigm. Instead, the paper proposes to question some possible reasons for conflicting dynamics that could stall positive progress for policy-making, and suggests an avenue for a higher order resolution. Most of all, the paper hopes to pragmatically encourage the exploration of opportunities for shared work and suggested that work on such opportunities, where it is found, can be well grounded through an incompletely theorized agreement. We invite scholars in the ethical AI community to explore the strengths and limits of this tool.

**Acknowledgements:** For valuable comments and feedback, the authors would like to thank Seth Baum, Haydn Belfield, Ashwin Acharya, John Danaher, Alex Lintz, Vincent Müller, Carina Prunkl, Rohin Shah and Jess Whittlestone, as well as two anonymous reviewers. Any remaining errors are all our own. No conflict of interest is identified.